\documentclass[aps, showpacs, showkeys,nofootinbib,floatfix]{revtex4}

\usepackage{amssymb}
\usepackage{amsmath}
\usepackage{graphicx}

\begin{document}

\title{Probing Ricci dark energy model with perturbations by using WMAP seven-year
cosmic microwave background measurements, BAO and Type Ia supernovae}

\author{Yuting Wang$^1$, Lixin Xu$^{2,}$$^{1,}$\footnote{lxxu@dlut.edu.cn}, and Yuanxing Gui$^1$}

\affiliation{$1$School of Physics and Optoelectronic Technology,
Dalian University of Technology, Dalian, Liaoning 116024, People's
Republic of China \\
$^2$ International Center for Astrophysics, Korea Astronomy and
Space Science Institute, Yuseong Daedeokdaero 776, Daejeon 305-348,
Republic of Korea}

\begin{abstract}
In this paper, we investigate the Ricci dark energy model with
perturbations through the joint constraints of current cosmological
data sets from dynamical and geometrical perspectives. We use the
full cosmic microwave background information from WMAP seven-year
data, the baryon acoustic oscillations from the Sloan Digital Sky Survey
and the Two Degree Galaxy Redshift Survey, and type Ia supernovae
from the Union2 compilation of the Supernova Cosmology Project
Collaboration. A global constraint is performed by employing the
Markov chain Monte Carlo method. With the best-fitting results, we
show the differences of cosmic microwave background power spectra
and background evolutions for the cosmological constant model and
Ricci dark energy model with perturbations.
\end{abstract}



\maketitle
\section{Introduction}
The present observations from WMAP seven-year cosmic microwave
background (CMB) radiation measurements \cite{ref:wmap7}, large-scale
structure surveys \cite{ref:Tegmark1,ref:Tegmark2}, and
updated type Ia supernovae (SNIa) \cite{ref:sn,ref:SN557} have
confirmed at a high confidence level the first implication from SNIa
in 1998 \cite{ref:Riess98,ref:Perlmuter99} that our Universe is
undergoing an accelerated expansion. It has been an important issue
of modern cosmology to unveil the mysterious face of the driving
force of current accelerated expansion. One of the theoretical
viewpoints is that dark energy with negative pressure drives current
expansion. A flood of dark energy models have been proposed, such as
a natural cosmological constant ($\Lambda$CDM model), dynamical dark
energy models with scalar fields \cite{ref:scal} or with exotic
equation of state \cite{ref:GCG, ref:holo}, and so on. Please see
\cite{ref:review} for an up-to-date review on dark energy.

Although the cosmological constant model keeps a good fit to the
current observations, it suffers from the fine-tuning problem and the cosmic
coincidence problem. In the final analysis, both are related to the
energy density of vacuum energy. Therefore, the cosmological
constant problem is in essence an issue of quantum gravity. Before
the complete theory of quantum gravity is established, the dark
energy model on the basis of the holographic principle of quantum
gravity theory was built in Ref. \cite{ref:holo} by applying
the energy bound proposed by Cohen, Kaplan, and Nelson \cite{ref:energybound} to
cosmology. Namely, for a system with size $L$ and UV cutoff
$\Lambda$ without decaying into a black hole, it is required that
the total energy in a region of size $L$ should not exceed the mass
of a black hole of the same size; thus $L^3\rho_\Lambda\leq
LM_{pl}^2$. It reveals a duality between the UV cutoff and IR cutoff.
In cosmology, the UV cutoff is related to the dark energy density,
and the IR cutoff is related to the large scale of the Universe. The
holographic dark energy models with different IR cutoffs have been
studied and the detailed discussions can be seen in Refs.
\cite{ref:holo,ref:holo2,ref:holo3,ref:holo4,ref:holo5}.

In Ref. \cite{ref:holo2}, Gao, Chen, and Shen first proposed that
dark energy density is in direct proportion to the Ricci scalar
curvature, $R$, and referred to this as the Ricci dark energy (RDE)
model. The distinct characteristic of the RDE model is that the
cosmic coincidence problem and causality problem can be naturally
solved. However, the physical motivation for such a model was vague
in Ref. \cite{ref:holo2}. Subsequently, Cai, Hu, and Zhang
\cite{ref:motivation} were inspired by the motivation of the causal
entropy bound in the framework of holography, found the causal
connection scale, $R_{CC}$, and put forward an appropriate physical
mechanism for this model. In \cite{ref:motivation}, it was pointed
out that the scale on which the black hole can be formed must be
within $R_{CC}$, which is set by the "Jeans" length of the
perturbations to $R_{CC}^{-2}=Max(\dot{H}+2H^2,-\dot{H})$, while one
assumes a black hole in the Universe is formed by gravitational
collapse of perturbations of cosmological spacetime. Only if one
takes the causal connection scale with $R_{CC}^{-2}=\dot{H}+2H^2$ as
the IR cutoff, can the dark energy model be consistent with current
observations \cite{ref:motivation}. It is found that
$R_{CC}^{-2}\propto$ $R$. Therefore, the RDE is stressed to be the
holographic dark energy with the IR cutoff of the causal connection
scale. The RDE density reads
\begin{equation}
\rho_{de}=3 M_{pl}^2
\alpha\left(\dot{H}+2H^2\right),\label{eq:rhoRDE}
\end{equation}
where $M_{pl}$ is the reduced Planck mass $M_{pl}\equiv1/\sqrt{8\pi
G}$, and $\alpha$ is the dimensionless parameter and has been
constrained by using the distance measurements from different joint
observations \cite{ref:previousRicci1,ref:previousRicci2}.
In particular, in these studies, the compressed CMB information is
used, including the "shift parameter," the "acoustic scale," and the
photon decoupling epoch, the values of which are obtained on the basis
of a given model in advance \cite{ref:lAR}. That is to say, before
these distance parameters are used to constrain a model, one should
renew to get their values basing on the constrained model. In this
paper, we revisit the cosmological parameters in the RDE model by using
the full CMB power spectrum data from WMAP7 and additional distance
measurements from baryon acoustic oscillations (BAOs) and Union2 SNIa
including 557 data points. Compared with the previous works, we not
only focus on the constraint of background evolution, but also take
dark energy perturbations into account. For there will be dark
energy perturbations according to the perturbed conservation
equations when the dark energy is not a cosmological constant.
What is more, the perturbations of dark energy play an important role
when the dark energy model is confronted with the current
observations \cite{ref:perrole}. The inclusion of dark energy
perturbations leaves an imprint on the CMB power spectrum
\cite{ref:0307100,ref:0307104} and leads to the changes of
parameters' spaces in the complete fitting to observational data
\cite{ref:0307104,ref:quintomfirst,ref:quintomused1,ref:quintomused2,ref:quintomused3,ref:quintomused4}.
In particular, there are the significant changes on the large-scale
CMB power spectra by the late integrated Sachs-Wolfe (ISW) effect
\cite{ref:ISW0}, which arises from a time-dependent gravitational
potential when dark energy dominates in the Universe at late time
for a flat dark energy model. The magnitude of the ISW-power spectrum is
dependent on dark energy models and dark energy perturbations
\cite{ref:0307100,ref:0307104,ref:ISW1,ref:ISW2,ref:ISW3,ref:ISW4,ref:ISW5,ref:ISW6}.
Therefore, a complete constraint on the RDE model with perturbations is
performed by employing the Markov chain Monte Carlo (MCMC) method.

The paper is organized as follows. In the next section, we review the
background equations for the RDE model and the perturbation equations in
the conformal Newtonian gauge. In Sec. III, we perform a global
fitting to observational data by using the MCMC method and discuss the
constraint results. The last section is the conclusion.

\section{Review of the dynamics of Ricci dark energy: Background and Perturbation evolution}
First of all, we give a brief review on the background evolution of the
RDE model in the framework of the unperturbed and spatially flat
Friedmann-Robertson-Walker metric. The Einstein field equation
is written as
\begin{eqnarray}
&&H^2=\frac{1}{3M_{pl}^2}(\rho_m+\rho_{de}), \label{eq:Fried}
\end{eqnarray}
where the energy density of RDE, $\rho_{de}$, is rewritten with the
variable $x=\ln a$ as
\begin{eqnarray}
\rho_{de}=3\alpha
M_{pl}^2\left(\frac{1}{2}\frac{dH^2}{dx}+2H^2\right).
\label{eq:rhoR}
\end{eqnarray}

After using the dimensionless definitions $E=\frac{H}{H_0}$ and
$\Omega_{m0}=\frac{\rho_{m0}}{3M_{pl}^2H_0^2},$ we can rewrite Eq.
(\ref{eq:Fried}) as

\begin{eqnarray}
&&E^2=\Omega_{m0}e^{-3x}+\alpha\left(\frac{1}{2}\frac{dE^2}{dx}+2E^2\right).
\end{eqnarray}
The solution of this first order differential equation with the
initial condition $E^2(x=0)=1$ is obtained:
\begin{eqnarray}
E^2=\frac{2}{2-\alpha}\Omega_{m0}e^{-3x}+\left(1-\frac{2}{2-\alpha}\Omega_{m0}\right)e^{-(4-\frac{2}{\alpha})x}.
\label{eq:E2}
\end{eqnarray}
From Eqs. (\ref{eq:rhoR},\ref{eq:E2}), we can get the dimensionless
energy density of RDE
\begin{eqnarray}
\Omega_{de}(x)=\frac{\rho_{de}}{3M_{pl}^2H_0^2}&&{=}\frac{\alpha}{2-\alpha}\Omega_{m0}e^{-3x}+\left(1-\frac{2}{2-\alpha}\Omega_{m0}\right)e^{-(4-\frac{2}{\alpha})x} \nonumber\\
&&{=}\frac{\alpha}{2-\alpha}\Omega_{m0}e^{-3x}+\left(\Omega_{de0}-\frac{\alpha}{2-\alpha}\Omega_{m0}\right)e^{-(4-\frac{2}{\alpha})x},
\end{eqnarray}
where $\Omega_{de0}$ is the present value of the dimensionless
energy density of RDE. Then the corresponding equation of state
(EOS) of RDE is obtained by the conservation equation
\begin{eqnarray}
w(x)&&{=}-1-\frac{1}{3}\frac{d\ln \Omega_{de}(x)}{dx} \nonumber\\
&&{=}\frac{(\Omega_{de0}-\frac{\alpha}{2-\alpha}\Omega_{m0})\frac{\alpha-2}{\alpha}e^{-(4-\frac{2}{\alpha})x}}{3\Omega_{de}(x)}.
\end{eqnarray}

When the EOS of dark energy is not constantly equal to -1, it is
necessary to consider dark energy perturbations. We work in the
conformal Newtonian gauge \footnote{Refer to the reviews of
cosmological perturbation theory
\cite{ref:perturbationreview1,ref:perturbationreview2,ref:perturbationreview3,ref:perturbationreview4,ref:perturbationreview5}for
the detailed gauge transformation.}, the perturbed metric being
\begin{eqnarray}
&&ds^2=-a(\eta)^2\{(1+2\psi)d\eta^2-(1-2\phi)\delta_{ij}dx^idx^j\}.
\end{eqnarray}
The perturbed conservation equations of dark energy perturbations
\cite{ref:perturbationreview3} read
\begin{eqnarray}
&&\delta'=-(1+w)(\theta-3\phi')-3\mathcal{H}(c_s^2-w)\delta, \\
&&\theta'=-\mathcal{H}(1-3w)\theta-\frac{w'}{1+w}\theta+\frac{c_s^2}{1+w}k^2\delta+k^2\psi,
\end{eqnarray}
where $\delta$ is the energy density perturbation, and $\theta$ is the velocity divergence perturbation
and is related to the velocity by $\theta=ik^j \upsilon_j$\cite{ref:perturbationreview3}. Where the primes
denote the derivatives with respect to conformal time $\eta$,
$\mathcal{H}$ is the conformal Hubble function and $c_s^2$ is the
general sound speed, being defined as
\begin{eqnarray}
&&c_s^2\equiv\frac{\delta p}{\delta \rho}.
\end{eqnarray}
From the intrinsic entropy perturbation $\Gamma$
\cite{ref:0307100,ref:perturbationreview2},
\begin{eqnarray}
&&w\Gamma\equiv(c_s^2-c_a^2)\delta=\frac{p'}{\rho}\left(\frac{\delta
p}{p'}-\frac{\delta \rho}{\rho'}\right), \label{eq:gamma}
\end{eqnarray}
where the adiabatic speed of sound, $c_a^2$, is purely determined by the
EOS,
\begin{eqnarray}
&&c_a^2\equiv\frac{p'}{\rho'}=w-\frac{w'}{3\mathcal{H}(1+w)},
\label{eq:ca}
\end{eqnarray}
it is found that only in the rest frame where $\delta$ keeps gauge
invariant can the general sound speed be a gauge invariant quantity
since $\Gamma$ and $c_a^2$ are gauge and scale independent.

The following transformation gives the useful relation between the
gauge invariant, rest frame density perturbation, $\hat{\delta}$,
the density and velocity divergence perturbations in a general frame,
$\delta$ and $\theta$ \cite{ref:0307100,ref:perturbationreview2}:
\begin{eqnarray}
&&\hat{\delta}=\delta+3\mathcal{H}(1+w)\frac{\theta}{k^2}.
\label{eq:delta}
\end{eqnarray}
Thus we can express the pressure perturbation in a general frame,
$\delta p$, in terms of the general sound speed in the rest frame,
$\hat{c}_s^2$, by using Eqs. (\ref{eq:gamma},\ref{eq:delta})
\begin{eqnarray}
&&\delta p=\hat{c}_s^2\delta
\rho+3\mathcal{H}(1+w)(\hat{c}_s^2-c_a^2)\rho\frac{\theta}{k^2}.
\end{eqnarray}
Using the above equation and Eq. (\ref{eq:ca}), we have
\begin{eqnarray}
&&\delta'=-(1+w)(\theta-3\phi')-3\mathcal{H}(\hat{c}_s^2-w)\delta-3\mathcal{H}(w'+3\mathcal{H}(1+w)(\hat{c}_s^2-w))\frac{\theta}{k^2}, \label{eq:delta2}\\
&&\theta'=-\mathcal{H}(1-3\hat{c}_s^2)\theta+k^2\left(\frac{\hat{c}_s^2\delta}{1+w}+\psi\right).
\label{eq:theta}
\end{eqnarray}

\section{Constraint Method and Results from date sets: WMAP, BAOS, and SNIA}
\subsection{Constraint Method and Data}
In the process of performing the observational constraints on
cosmological parameters, we adopt the MCMC method by working with
the publicly available {\bf CosmoMC} package \cite{ref:MCMCweb,ref:MCMC},
including the {\bf CAMB} \cite{ref:cambweb} code of
calculating the theoretical CMB power spectrum. We modified the code
for the RDE model with dark energy perturbations. In the first instance,
work including perturbations of dark energy with EOS on
either side of the cosmological constant boundary, i.e., $w>-1$ or
$w<-1$, was studied \cite{ref:noqunitom}. In this case, the
perturbed equations of dark energy can be solved well. However, when
one encounters the perturbations of dark energy with EOS crossing
the cosmological constant boundary -1, there inevitably exists a
divergent problem due to the term $1+w$ in the denominator in
Eq.(\ref{eq:theta}). As pointed out in \cite{ref:previousRicci2},
RDE behaves like a "quintom" when $\alpha<\frac{1}{2}$. Here, we use
the method proposed in \cite{ref:quintomfirst}, dividing the whole
range of $w$ into three regions with a small positive parameter
$\epsilon$: (1) $w>-1+\epsilon$; (2) $-1+\epsilon\geq
w\geq-1-\epsilon$; and (3) $w<-1-\epsilon$. Equations
(\ref{eq:delta2},\ref{eq:theta}) are well satisfied in regions
(1) and (3). The divergence happens in region (2). In order to
deal with the divergence and keep dark energy perturbations
continuous, the perturbations in region (2) are set as follows
\cite{ref:quintomfirst}:
\begin{eqnarray}
&&\delta'=0, ~~~~\theta'=0.
\end{eqnarray}
While in the process of running the program, we set
$\epsilon=10^{-7}$, which is within the limited range,
$\epsilon<10^{-5}$\cite{ref:quintomfirst}. By using the method,
Refs.
\cite{ref:quintomused1,ref:quintomused2,ref:quintomused3,ref:quintomused4}
gave the global fittings on the quintomlike dark energy with
parametrized EOS when the dark energy perturbations were included.

We adopt the following seven-dimensional cosmological spaces:
\begin{eqnarray}
&&P\equiv(\omega_b, \omega_c, \Theta_S, \tau, \alpha, n_s,
\log[10^{10}A_s])
\end{eqnarray}
where $\omega_b=\Omega_bh^2$ and $\omega_c=\Omega_ch^2$ are the
physical baryon and cold dark matter densities, $\Theta_S$ is the
ratio (multiplied by 100) of the sound horizon and angular diameter
distance, $\tau$ is the optical depth, $\alpha$ is the newly added
Ricci parameter, $n_s$ is the scalar spectral index, and $A_s$ is
defined as the amplitude of the initial power spectrum. The pivot scale
of the initial scalar power spectrum we have used is $k_{s0}=0.05
\textmd{Mpc}^{-1}$. The corresponding priors are taken as follows:
$\Omega_{b}h^2\in[0.005,0.1]$, $\Omega_{c}h^2\in[0.01,0.99]$,
$\Theta_S\in[0.5,10]$, $\tau\in[0.01,0.8]$, $\alpha\in[0.1,0.8]$,
$n_s\in[0.5,1.5]$, and $\log[10^{10}A_s]\in[2.7,4]$. In addition, we
use a top-hat prior of the cosmic age, i.e., $10 \textmd{Gyr}<t_0<20
\textmd{Gyr}$, impose a weak Gaussian prior on the physical baryon
density $\Omega_{b}h^2=0.022\pm0.002$ from big bang nucleosynthesis
\cite{ref:bbn}, and use the new value of the Hubble constant
$H_0=74.2\pm3.6 {\rm km ~s}^{-1} {\rm Mpc}^{-1}$ \cite{ref:0905} by
a Gaussian likelihood function.

In our calculations, we have taken the total likelihood $L\propto
e^{-\chi^2/2}$ to be the products of the separate likelihoods of
CMB, BAOs, and SNIa. Thus we have
\begin{eqnarray}
\chi^2=\chi^2_{CMB}+\chi^2_{BAO}+\chi^2_{SNIa}.
\end{eqnarray}
The full CMB data are used, including the new temperature and
polarization power spectra from WMAP seven-year
data \cite{wmapdata7}. We
combine the additional distance measurements from BAOs and SNIa. For
BAO information, we use the values of $[r_s(z_d)/D_V(0.2),
r_s(z_d)/D_V(0.35)]$ and their inverse covariance matrix
\cite{ref:Percival2}. For SNIa, we use the 557 Union2 SNIa data with
systematic errors \cite{ref:SN557}. In order to assess the goodness
of fit between the RDE model and $\Lambda$CDM model on the basis of the
same observational data, we also perform the best fitting to the
WMAP+BAO+SNIa data sets in the $\Lambda$CDM model.
\subsection{Fitting Results and Discussions}
In this section, we first present the constraint results from
observational data. Then using the fitting results, we investigate
the imprints on CMB temperature anisotropy, background evolution,
and evolutions of perturbations for the RDE model with perturbations.

\begin{figure}[!htbp]
\includegraphics[width=20cm,height=14cm]{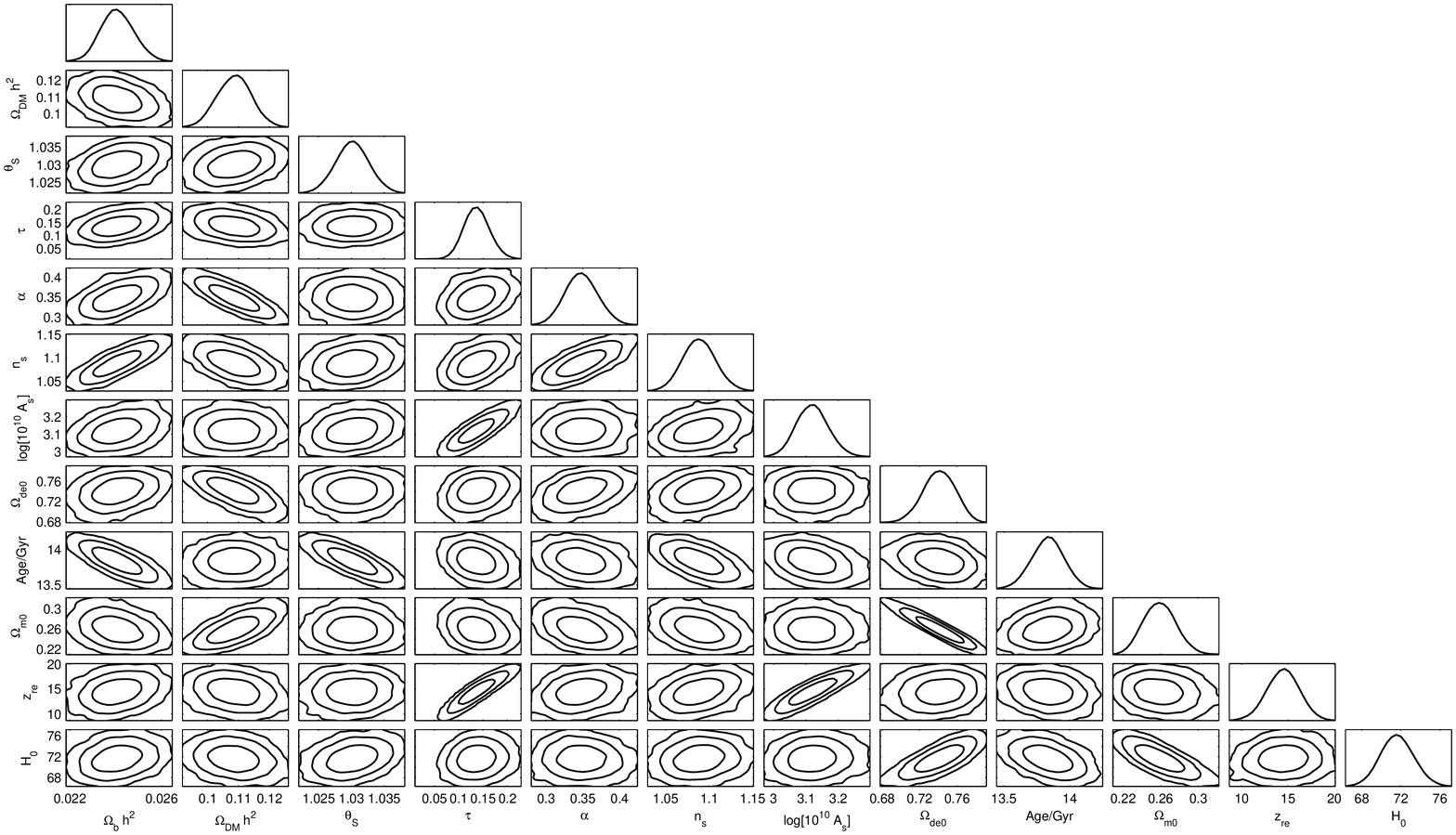}
  \caption{The 1D marginalized distributions on individual parameters and 2D contours with 68\%C.L., 95 \%C.L., and 99.7\%C.L.
   between each other using the combination of the observational
  data from WMAP, BAOs and SNIa for the RDE model.
 }\label{fig:all}
\end{figure}

\begin{figure}[!htbp]
\includegraphics[width=13cm,height=9cm]{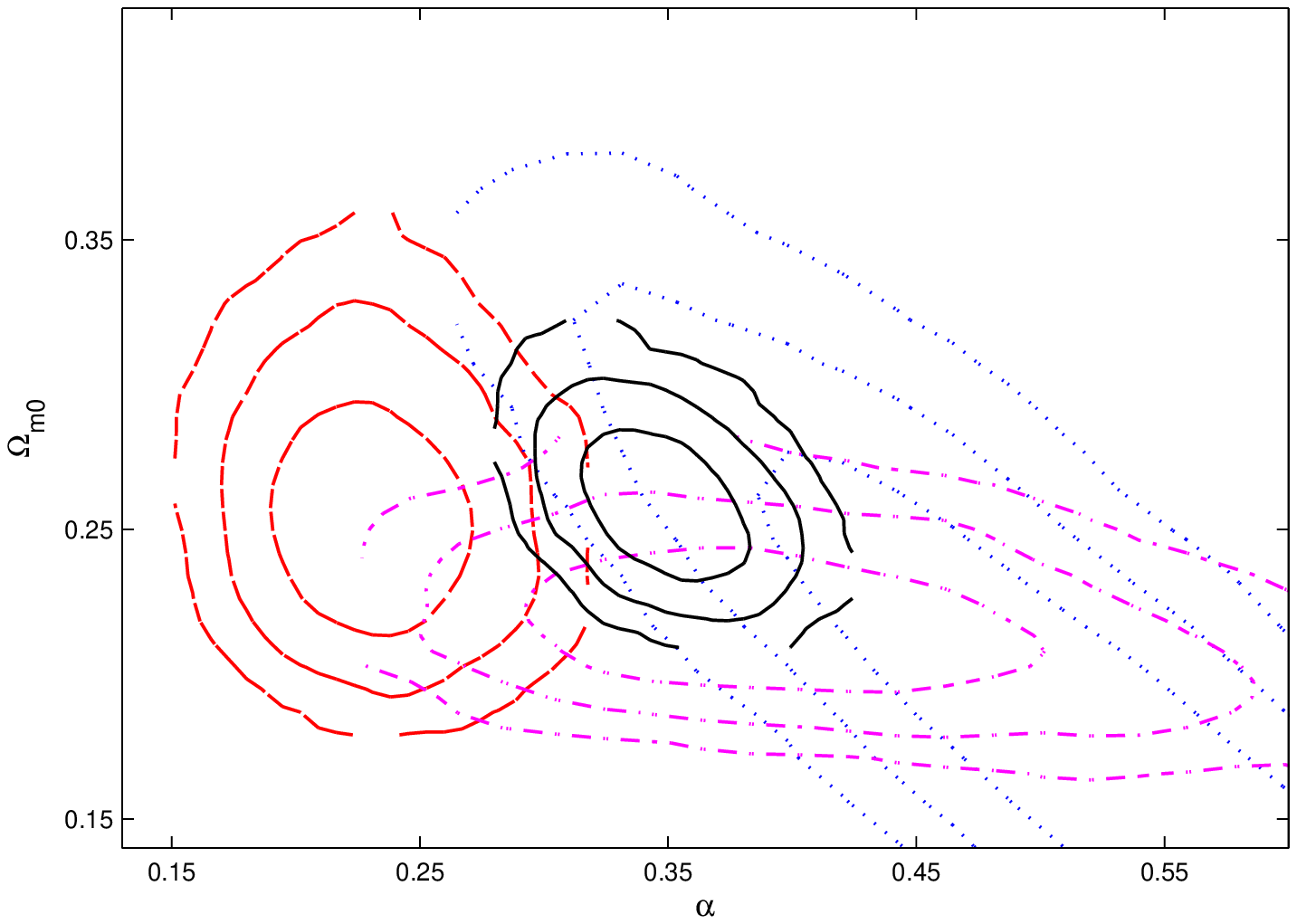}
  \caption{The 2D contours with 68\%C.L., 95 \%C.L., and 99.7\%C.L.
  between $\alpha$ and $\Omega_{m0}$ for the RDE model
  from WMAP alone (red dashed line), BAOs alone (magenta dash-dotted
  line), SNIa alone (blue dotted line), and the combination of WMAP,
  BAOs and SNIa (solid black line).
  }\label{fig:alpha_omega}
\end{figure}

The constraint results of full basic parameters and derived
parameters are showed in Table \ref{tab:results}, where we list the
mean values with $1\sigma, 2\sigma$ regions and best-fit values from
WMAP alone in the RDE model, WMAP+BAO+SNIa in the RDE model, and
WMAP+BAO+SNIa in the $\Lambda$CDM model, respectively. It is seen
that the inclusion of distance information from BAOs and SNIa leads
to a significant change for best-fit value of parameter $\alpha$.
Both the best-fit values of the scalar spectral index, $n_s$, in the
RDE model are greater than 1. That is to say, the scalar spectrum is
"blue" tilted. This result is different from that in the
$\Lambda$CDM model, where within $2\sigma$ regions the initial power
spectrum with a blue tilt is not favored. For the combined
constraint, the total best-fit $\chi^2$ of the RDE model is 8056.1,
while $\chi^2=8002.6$ for the best-fit $\Lambda$CDM model.
Therefore, the $\Lambda$CDM model still shows a better fit to the
current data than the RDE model. In Fig. \ref{fig:all}, we give the
one-dimensional (1D) marginalized distributions of parameters and 2D contours with
confidence level (C.L.) from the combination of WMAP, BAOs, and SNIa.
The respective contributions of WMAP, BAOs, and SNIa to cosmological
parameters can be seen in Fig. \ref{fig:alpha_omega}, which shows
the 2D contours of parameters $(\alpha, \Omega_{m0})$ from
WMAP alone, BAOs alone, SNIa alone, and their combinations. The
stringent results can be obtained from the joint constraints of
dynamical and geometrical aspects.
\begin{table}
\begin{center}
\begin{tabular}{|cc|   cc|   cc|   cc| }
\hline \hline Model and data &&RDE and WMAP alone& & RDE and WMAP+BAO+SNIa &
& $\Lambda$CDM and WMAP+BAO+SNIa&
\\ \hline
Parameters &&~~~~~~~~~~~~Mean~~~~~~~~~~~~~~Best fit & &
~~~~~~~~~~~~Mean~~~~~~~~~~~~~~~Best fit & &
~~~~~~~~~~~~Mean~~~~~~~~~~~~~~~Best fit &
\\ \hline
$\Omega_{b}h^2$ &&$0.0222^{+0.0006
+0.0012}_{-0.0006-0.0011}$~~~~~~~$0.0222$ & &$0.0241^{+0.0007
+0.0014}_{-0.0007-0.0013}$~~~~~~~~~$0.0241$ & &$0.0225^{+0.0005
+0.0010}_{-0.0005-0.0010}$~~~~~~~~~$0.0226$ &\\
$\Omega_{c}h^2$ &&$0.1289^{+0.0064
+0.0121}_{-0.0063-0.0125}$~~~~~~~$0.1313$ & &$0.1087^{+0.0052
+0.0104}_{-0.0055-0.0103}$~~~~~~~~~$0.1086$ &&$0.1124^{+0.0033
+0.0065}_{-0.0033-0.0063}$~~~~~~~~~$0.1123$ &\\
$\Theta_S$ &&$1.0296^{+0.0024
+0.0047}_{-0.0023-0.0048}$~~~~~~~$1.0293$ & &$1.0302^{+0.0024
+0.0048}_{-0.0025-0.0048}$~~~~~~~~~$1.0303$ &&$1.0392^{+0.0026
+0.0051}_{-0.0025-0.0050}$~~~~~~~~~$1.0393$ &\\
$\tau$ &&$0.0986^{+0.0077 +0.0309}_{-0.0092-0.0285}$~~~~~~~$0.0931$
& &$0.1374^{+0.0107 +0.0417}_{-0.0122-0.0385}$~~~~~~~~~$0.1382$
&&$0.0877^{+0.0063 +0.0245}_{-0.0070-0.0230}$~~~~~~~~~$0.0877$
&\\
$\alpha$ &&$0.2310^{+0.0252
+0.0526}_{-0.0257-0.0470}$~~~~~~~$0.2180$ & &$0.3484^{+0.0217
+0.0433}_{-0.0215-0.0407}$~~~~~~~~~$0.3452$ &&-~~~~~~~~~~~~~~~~~~~~~~~~~~~~~~~~~~- &\\
$n_s$ &&$1.0159^{+0.0186 +0.0385}_{-0.0186-0.0355}$~~~~~~~$1.0080$ &
&$1.0887^{+0.0181 +0.0374}_{-0.0185-0.0351}$~~~~~~~~~$1.0871$
&&$0.9680^{+0.0119
+0.0238}_{-0.0119-0.0235}$~~~~~~~~~$0.9691$ &\\
$\log[10^{10}A_s]$ &&$3.113^{+0.036
+0.073}_{-0.037-0.072}$~~~~~~~~~~$3.109$ & &$3.120^{+0.048
+0.100}_{-0.047-0.090}$~~~~~~~~~~~~$3.122$ &&$3.187^{+0.036
+0.072}_{-0.037-0.071}$~~~~~~~~~~~~$3.180$ &\\
\hline $\Omega_{de0}$ &&$0.744^{+0.026
+0.047}_{-0.025-0.056}$~~~~~~~~~~$0.745$ & &$0.741^{+0.017
+0.031}_{-0.017-0.034}$~~~~~~~~~~~~$0.746$ &&$0.727^{+0.014
+0.027}_{-0.015-0.03}$~~~~~~~~~~~~~$0.728$ &\\
$Age/Gyr$ &&$14.00^{+0.12 +0.23}_{-0.11-0.22}$~~~~~~~~~~~~~$14.03$ &
&$13.83^{+0.12 +0.23}_{-0.12-0.23}$~~~~~~~~~~~~~~~$13.82$
&&$13.76^{+0.011
+0.021}_{-0.011-0.021}$~~~~~~~~~~~~~$13.75$ &\\
$\Omega_{m0}$ &&$0.256^{+0.025
+0.056}_{-0.026-0.047}$~~~~~~~~~~$0.255$ & &$0.259^{+0.017
+0.035}_{-0.017-0.031}$~~~~~~~~~~~~$0.254$ &&$0.273^{+0.015
+0.03}_{-0.014-0.027}$~~~~~~~~~~~~~$0.272$ &\\
$z_{re}$ &&$12.36^{+1.50 +2.99}_{-1.50-2.96}$~~~~~~~~~~~~~$12.02$ &
&$14.49^{+1.68 +3.28}_{-1.69-3.41}$~~~~~~~~~~~~~~~$14.61$
&&$10.51^{+1.18 +2.34}_{-1.17-2.35}$~~~~~~~~~~~~~~$10.32$
&\\
$H_0$ &&$77.13^{+3.35 +6.56}_{-3.31-6.57}$~~~~~~~~~~~~~$77.54$ &
&$71.63^{+1.66 +3.30}_{-1.63-3.21}$~~~~~~~~~~~~~~~$72.26$
&&$70.32^{+1.28+2.52}_{-1.27-2.47}$~~~~~~~~~~~~~~~$70.38$ &\\
\hline \hline
\end{tabular}
\caption{The mean values with $1\sigma$, $2\sigma$ regions and best-fit
values for the RDE model from WMAP alone and the combination of
WMAP, BAOs, and SNIa, respectively, and the mean values with
$1\sigma$, $2\sigma$ regions and best-fit values for the $\Lambda$CDM
model from the combination of WMAP, BAOs and
SNIa.}\label{tab:results}
\end{center}
\end{table}

\begin{figure}[ht]
\includegraphics[width=13cm,height=9cm]{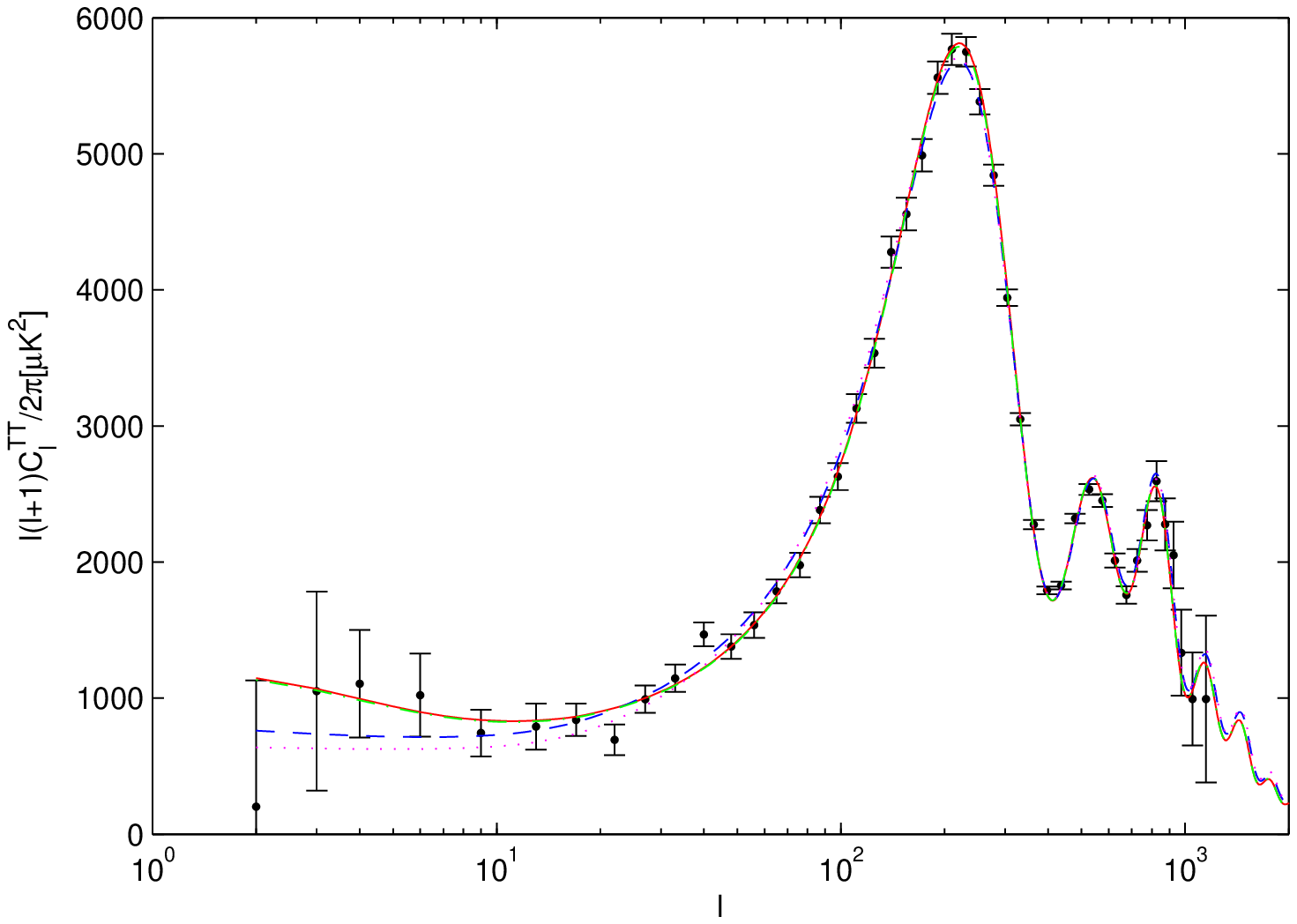}
  \caption{CMB temperature power spectra vs multipole moment $l$, where the black dots with black error bars denote the
  observed data with their corresponding uncertainties from WMAP, the red solid line is for the $\Lambda$CDM model with the
  best-fit values from the WMAP alone constraint \cite{ref:wmap7}, the green dash-dotted line is for the $\Lambda$CDM model with
  the best-fit values from the WMAP+BAO+SNIa joint constraint, the blue dashed line is for RDE with the best-fit values from the WMAP alone
  constraint, and the magenta dotted line is for the RDE model with the best-fit values from the WMAP+BAO+SNIa joint
  constraint.
 }\label{fig:cl}
\end{figure}

\begin{figure}[!htbp]
\includegraphics[width=13cm,height=9cm]{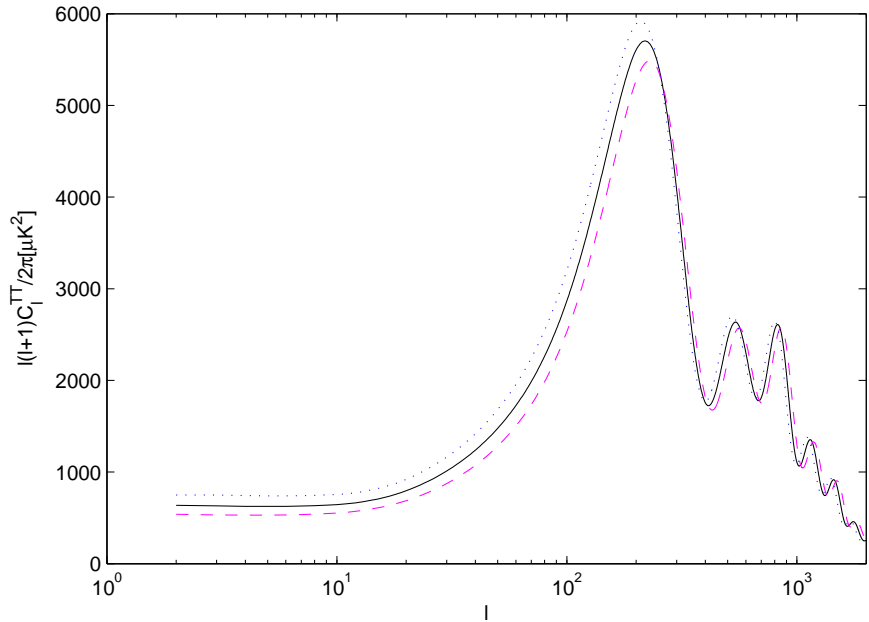}
  \caption{CMB temperature power spectra vs multipole moment $l$
  for the RDE model with different parameters, $\alpha$,
  where the blue dotted line is the power spectrum with $\alpha=0.44$, the black solid
  line is the power spectrum with the best-fit value of $\alpha$ from the WMAP+BAO+SNIa joint
  constraint, and the magenta dashed line is the power spectrum with $\alpha=0.24$.
 }\label{fig:cl_alphadiff}
\end{figure}

\begin{figure}[!htbp]
\includegraphics[width=10.3cm,height=8.5cm]{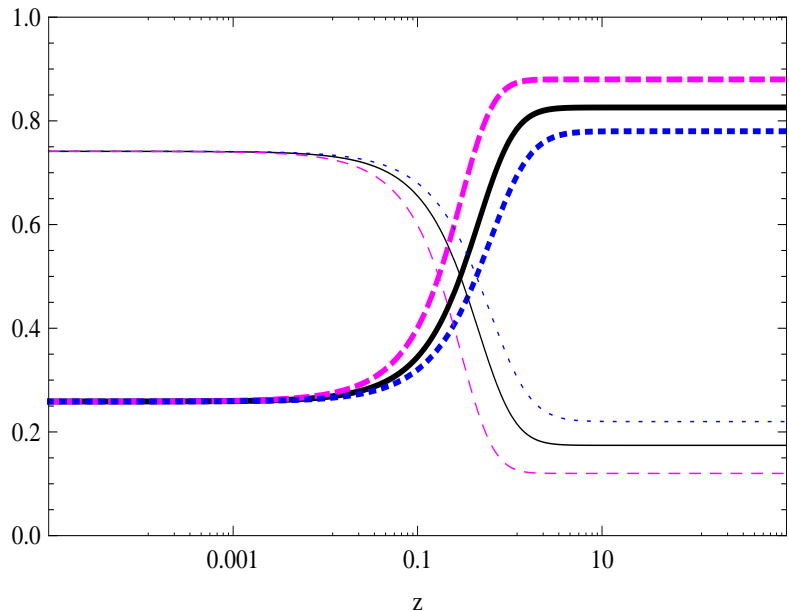}
  \caption{The usual dimensionless energy
  densities of RDE (thin) $\left(\frac{\Omega_{de}}{E^2}\right)$ and matter
  component (thick)
  $\left(\frac{\Omega_{m0}(1+z)^3}{E^2}\right)$
  vs redshift $z$ for the RDE model with different parameters,
  $\alpha$, where the blue dotted lines are dimensionless energy
  densities with $\alpha=0.44$, the black solid
  lines are dimensionless energy
  densities with the best-fit value of $\alpha$ from the WMAP+BAO+SNIa joint
  constraint, and the magenta dashed lines are dimensionless energy
  densities with $\alpha=0.24$.
 }\label{fig:omegam_R}
\end{figure}

\begin{figure}[!htbp]
\includegraphics[width=13cm,height=9cm]{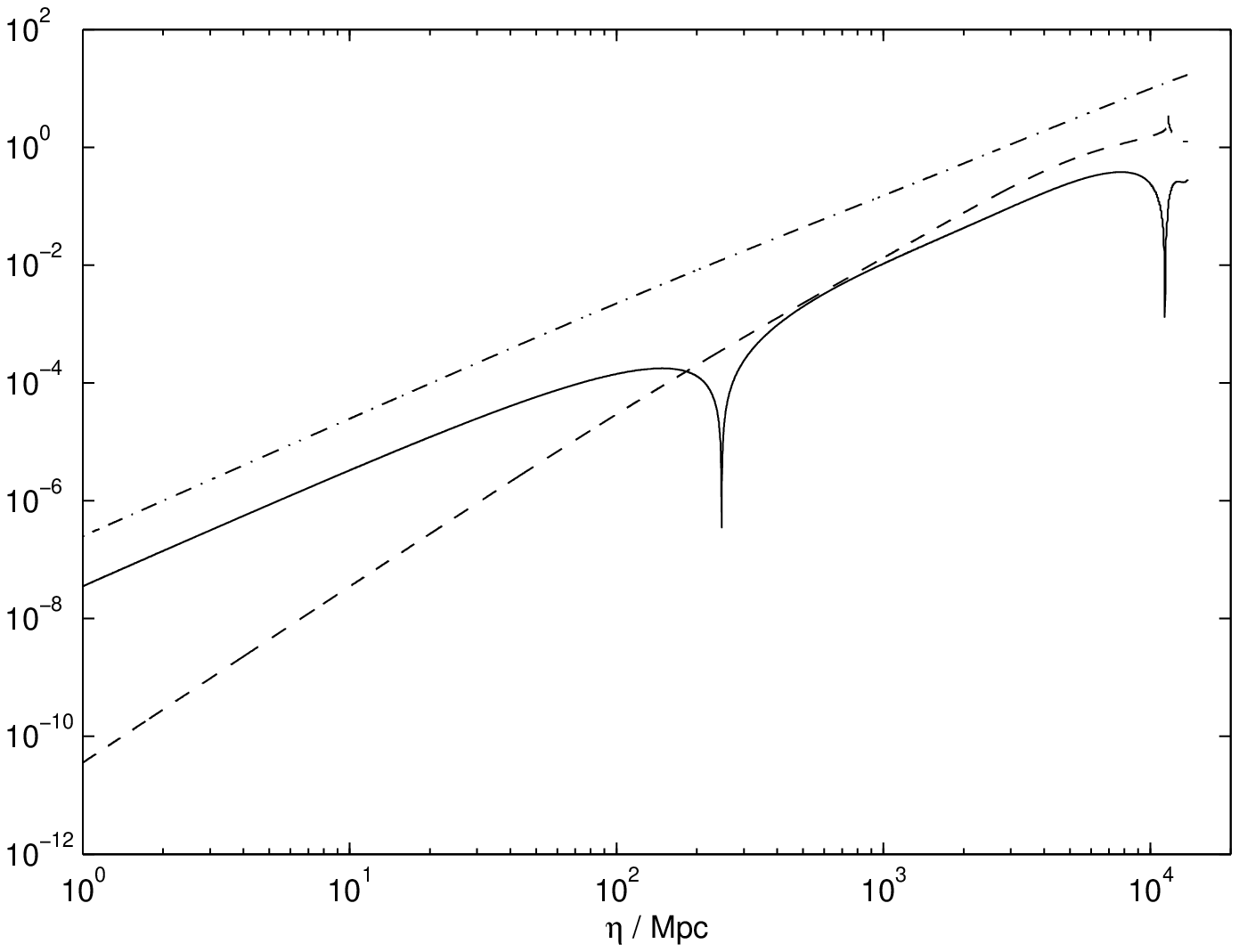}
  \caption{ The evolutions of cold dark matter density perturbation,
  $\delta_c$ (dash-dotted line), dark energy density perturbation, $|\delta_{de}|$ (solid line), and its velocity perturbation,
  $\upsilon_{de}$ (dashed line).
 }\label{fig:val_evol}
\end{figure}

In Fig. \ref{fig:cl}, we show CMB temperature power spectra for the
RDE model by using the best-fitting results in Table
\ref{tab:results} and for the $\Lambda$CDM model with the best-fit
values from the WMAP alone constraint \cite{ref:wmap7} and WMAP+BAO+SNIa
in Table \ref{tab:results}. It is found that there are significant
discrepancies of CMB temperature power spectra in the two models on
large angular scales, where the contribution to the secondary CMB
temperature power spectrum is mainly from the late ISW effect. Both
the curves of the CMB temperature power spectra for the RDE model on
large angular scales almost tend to be plat. A large-scale ($l<10$)
plateau happens in the RDE model. In order to understand the imprint
on the CMB temperature power spectrum from parameter $\alpha$ alone, we
illustrate the CMB temperature power spectra for different $\alpha$
in Fig. \ref{fig:cl_alphadiff}. It is shown that the CMB temperature
power spectra on all the scales decrease as the value of $\alpha$
becomes smaller. Because $\alpha$ has an effect on the expansion
rate, the angular diameter distance to recombination becomes larger
when $\alpha$ decreases. So there is the rightward shift of the
positions of peaks. What is more, the usual dimensionless densities
of the RDE and matter component are changed for different parameters,
$\alpha$, as shown in Fig. \ref{fig:omegam_R}. That is to say, the
change in the expansion rate brings about the differences of usual
dimensionless densities when there is the same matter density,
$\Omega_{m0}$. We can see at early times the usual dimensionless
matter density decreases when $\alpha$ become larger. We try to look
into the role of the change of $\alpha$ by taking another
point of view. We consider that this effect can be viewed as coming from the
different effective matter densities based on the same expansion
rate, i.e., $\Omega^{eff}_{m0}(1+z)^3/E^2_s$. So there is a lower
effective matter density with the increase in $\alpha$. The lower
effective matter density causes the delay of the epoch of
matter-radiation equality, i.e., the smaller $z_{eq}$. This change in
redshift of equality leads to the enhanced late ISW effect on large
scales and power spectra on small scales. Furthermore, the delay of
the time of dark energy domination for smaller $\alpha$ also leads
to the suppressed late ISW effect on large scales. In Fig.
\ref{fig:val_evol}, we show the evolutions of cold dark matter and
dark energy perturbations on the large scale, i.e.,
$k=0.001\textmd{Mpc}^{-1}$.


\begin{figure}[!ht]
\includegraphics[width=18cm,height=8cm]{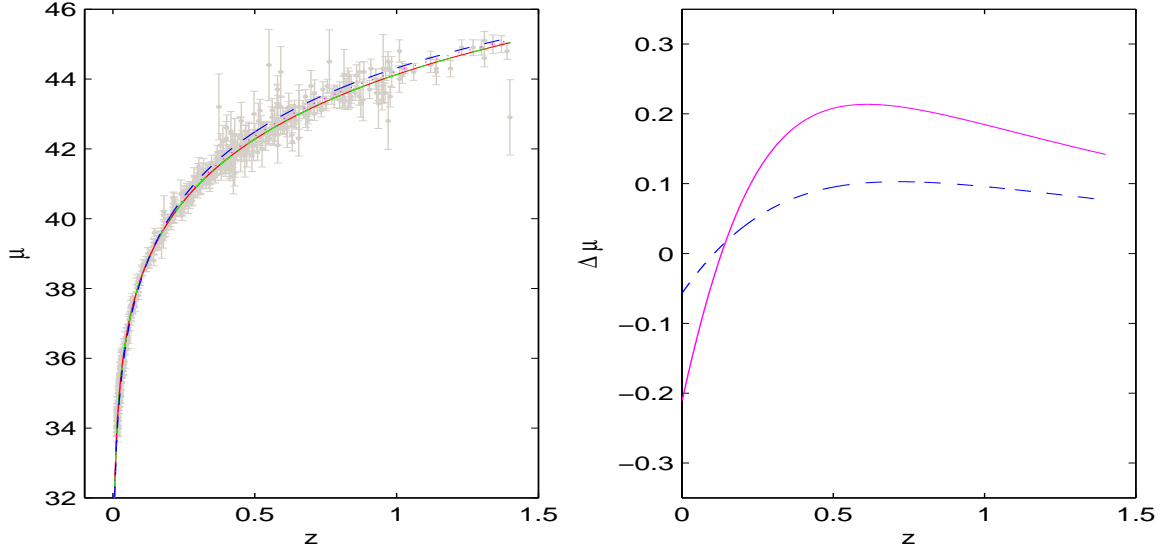}
  \caption{ Left panel: The distance moduli $\mu$ vs redshift $z$, where the grey dots with grey error bars denote the
  observed data with their corresponding uncertainties from SNIa, the red solid line is for the $\Lambda$CDM model with the
  best-fit values from the WMAP alone constraint \cite{ref:wmap7}, the green dash-dotted line is for the $\Lambda$CDM model
  with the best-fit values from the WMAP+BAO+SNIa joint constraint, the blue dashed line is for the RDE model with the best-fit values from the WMAP alone
  constraint, and the magenta dotted line is for the RDE model with the best-fit values from the WMAP+BAO+SNIa joint
  constraint. Right panel: The relative distance moduli $\Delta \mu =\mu_{\textmd{RDE}}-\mu_{\Lambda \textmd{CDM}}$ vs redshift $z$,
  where the magenta solid line is for the best-fit values from the WMAP alone
  constraint, and the blue dashed line is for the best-fit values from the WMAP+BAO+SNIa joint
  constraint.
 }\label{fig:mu2_z}
\end{figure}

\begin{figure}[!htbp]
\includegraphics[width=13cm,height=9cm]{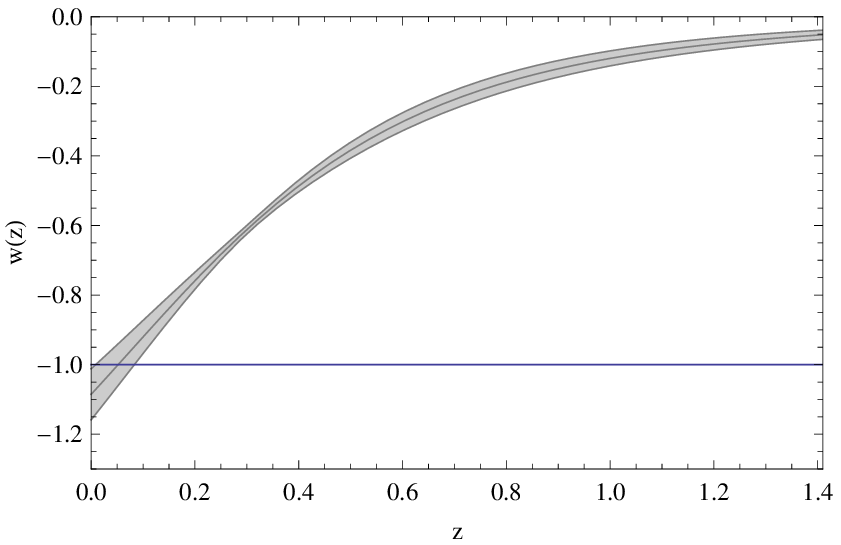}
  \caption{ The evolution of the equation of state with $1\sigma$
  errors, where the mean values with $1\sigma$ errors for the RDE model from the WMAP+BAO+SNIa joint
  constraint are used.
 }\label{fig:w_z}
\end{figure}

The differences of background evolutions between the $\Lambda$CDM
model and RDE model can be seen through the evolutions of the
distance moduli and relative distance moduli in Fig.
\ref{fig:mu2_z}. Then following the method of the propagation of
errors by the Fisher matrix analysis
\cite{ref:werror1,ref:werror2,ref:werror3}\footnote{ We refer to
\cite{ref:margnalized} to marginalize the cosmological parameters
which are not associated with the EOS and obtain the error covariance
matrix of EOS parameters from the sub-Fisher matrix.}, we plot the
evolution of EOS with $1\sigma$ errors in Fig. \ref{fig:w_z} by
using the mean values from WMAP+BAO+SNIa. The current value of the EOS
is -1.0853, less than -1. What is more, it is seen that the EOS of
RDE crosses the cosmological constant boundary at $z\sim0.05$.

\section{Summary}
In summary, we perform a global fitting on the RDE model with
perturbations to the combined constraints from the full CMB data,
BAOs, and SNIa by using the MCMC method. All the cosmological
parameters for the RDE model are well determined, as shown in Table
\ref{tab:results}. It is worth noting that the initial power
spectrum with a blue tilt is favored for the RDE model. Making use of
the constraint results, we investigate the CMB temperature power
spectra. It is found that the differences of CMB temperature power
spectra on large angular scales are significant between the
$\Lambda$CDM model and the RDE model with perturbations. The main
feature of CMB temperature power spectra on large angular scales for
the RDE model with perturbations is that there exists a flat
plateau. In view of the differences of large-scale power spectra, on
which the ISW effect plays an important role, the effects on RDE
parameters made by the ISW data from the cross correlations of CMB
andlarge-scale structure will be worth investigating in future work.

\acknowledgements{This work is supported by NSF (10703001) of People's Republic
of China and the Fundamental Research Funds for the Central
Universities (DUT10LK31).}


\begin{thebibliography}{*}
\bibitem{ref:wmap7} E. Komatsu {\it et al.}, Astrophys. J. Suppl. {\bf 192}, 18 (2011), arXiv:1001.4538.

\bibitem{ref:Tegmark1} M. Tegmark {\it et al.}, Phys. Rev. D {\bf 69}, 103501 (2004), astro-ph/0310723.

\bibitem{ref:Tegmark2} M. Tegmark {\it et al.}, Astrophys. J. {\bf 606}, 702 (2004), astro-ph/0310725.

\bibitem{ref:sn} A. Clocchiatti {\it et al.}, Astrophys. J. \textbf{642}, 1
(2006).

\bibitem{ref:SN557} R. Amanullah {\it et al.}, [Supernova Cosmology Project Collaboration], Astrophys. J. {\bf 716}, 712 (2010), arXiv:1004.1711.
\bibitem{ref:Riess98} A. G. Riess {\it et al.}, Astron. J. \textbf{116}, 1009 (1998), astro-ph/9805201.

\bibitem{ref:Perlmuter99} S. Perlmutter {\it et al.}, Astrophys. J. \textbf{517}, 565 (1999), astro-ph/9812133.


\bibitem{ref:scal}
  B. Ratra and P. J. E. Peebles, Phys. Rev. D. \textbf{37}, 3406
  (1988);

  M. S. Turner and M. White, Phys. Rev. D \textbf{56}, 4439 (1997);

  R. R. Caldwell, R. Dave, and P. J. Steinhardt, Phys. Rev. Lett. \textbf{80}, 1582
  (1998);

  P. J. Steinhardt, M. L. Wang, and I. Zlatev, Phys. Rev. D \textbf{59}, 123504
  (1999);


  R. R. Caldwell, Phys. Lett. B \textbf{545}, 23 (2002);

  N. N. Weinberg, Phys. Rev. Lett. \textbf{91}, 071301 (2003);

  S. Nojiri and S. D. Odintsov, Phys. Rev. D \textbf{72}, 023003
  (2005);

  B. Feng, X. L. Wang, and X. M. Zhang, Phys. Lett. B \textbf{607}, 35
  (2005);

  Z. K. Guo, Y. S. Piao, X. N. Zhang, and Y. Z. Zhang, Phys. Lett. B \textbf{608}, 177
  (2005).

\bibitem{ref:GCG} A. Y. Kamenshchik, U. Moschella, and V. Pasquier, Phys. Lett. B \textbf{511}, 265
(2001);

   M. C. Bento, O. Bertolami, and A. A. Sen, Phys. Rev. D \textbf{66}, 043507
   (2002);

   M. C. Bento, O. Bertolami, M. J. Reboucas, and P. T. Silva, Phys. Rev. D \textbf{73}, 043504 (2006).

\bibitem{ref:holo} M. Li, Phys. Lett. B \textbf{603}, 1 (2004).

\bibitem{ref:review} M. Li, X. D. Li, S. Wang, and Y. Wang,
arXiv:1103.5870.

\bibitem{ref:energybound} A. G. Cohen, D. B. Kaplan, and A. E.
Nelson, Phys. Rev. Lett. {\bf 82}, 4971 (1999), hep-th/9803132.


\bibitem{ref:holo2} C. Gao, X. Chen, and Y. G. Shen, Phys. Rev. D {\bf 79}, 043511 (2009), arXiv:0712.1394.


\bibitem{ref:holo3} L. N. Granda and A. Oliveros, Phys. Lett. B {\bf 669}, 275 (2008),
arXiv:0810.3149;

                    Y. T. Wang and L. X. Xu, arXiv:1004.3340.
\bibitem{ref:holo4} L. X. Xu, J. B. Lu, and W. B. Li, Eur. Phys. J. C {\bf 64}, 89
(2009).

\bibitem{ref:holo5} R.-J. Yang, Z.-H. Zhu, and F. Wu, Int. J. Mod. Phys. A {\bf 26}, 317
(2011).
\bibitem{ref:motivation} R. G. Cai, B. Hu and Y. Zhang, Commun. Theor. Phys. {\bf 51}, 954
(2009).

\bibitem{ref:previousRicci1}
L. Xu, W. Li, J. Lu, and B. Chang, Mod. Phys. Lett. A {\bf 24}, 1355
(2009);

                             M. Li, X. Li, and X. Zhang, arXiv:0912.3988;

                             M. Li, X. D. Li, S. Wang, and X. Zhang, JCAP {\bf 0906}, 036
                             (2009);

                             L. X. Xu and Y. T. Wang, JCAP {\bf 06}, 002
                             (2010).
 \bibitem{ref:previousRicci2}   X. Zhang, Phys. Rev. D {\bf 79}, 103509 (2009);
\bibitem{ref:lAR}  H. Li, J. Q. Xia, G. B. Zhao, Z. H. Fan, and X. M. Zhang, Astrophys. J. {\bf 683}, L1-L4 (2008).

\bibitem{ref:perrole} C. G. Park, J. C. Hwang, J. H. Lee, and H. Noh, Phys. Rev. Lett. {\bf 103}, 151303 (2009).

\bibitem{ref:0307100} R. Bean and O. Dore, Phys. Rev. D {\bf 69}, 083503 (2004), astro-ph/0307100.

\bibitem{ref:0307104} J. Weller and A. M. Lewis, Mon. Not. Roy. Astron. Soc. {\bf 346}, 987 (2003).

\bibitem{ref:quintomfirst} G. B. Zhao, J. Q. Xia, M. Li, B. Feng, and Xinmin Zhang, Phys.
Rev. D {\bf 72}, 123515 (2005), astro-ph/0507482.

\bibitem{ref:quintomused1}
 J. Q. Xia, G. B. Zhao, B. Feng, H. Li, and Xinmin Zhang, Phys. Rev. D {\bf 73}, 063521 (2006),
 astro-ph/0511625.
\bibitem{ref:quintomused2}
 G. B. Zhao, J. Q. Xia, B. Feng, and Xinmin Zhang, Int. J.
Mod. Phys. D {\bf 16}, 1229  (2007), astro-ph/0603621.
\bibitem{ref:quintomused3}
 J. Q. Xia, Y. F. Cai, T. T. Qiu, G. B. Zhao, and Xinmin Zhang, Int. J.
Mod. Phys. D {\bf 17}, 1229 (2008), astro-ph/0703202.
\bibitem{ref:quintomused4}
 J. Q. Xia, H. Li, G. B. Zhao, and Xinmin Zhang, Phys. Rev. D {\bf 78},
 083524 (2008), arXiv:0807.3878;
\bibitem{ref:ISW0}  R. K. Sachs and A. M. Wolfe, Astrophys. J. \textbf{147}, 73 (1967).


\bibitem{ref:ISW1}  B. M. Schaefer, Mon. Not. Roy. Astron. Soc. {\bf 388}, 1403 (2008), arXiv:0803.2239.

\bibitem{ref:ISW2}  Y. T. Wang, Y. X. Gui, L. X. Xu, and J. B. Lu, Phys. Rev. D \textbf{81}, 083514 (2010).

\bibitem{ref:ISW3}  G. Olivares, F. A. Brandela, and D. Pavon, Phys. Rev. D \textbf{77}, 103520 (2008).

\bibitem{ref:ISW4}  J. B. Dent, S. Dutta, and T. J. Weiler, Phys. Rev. D \textbf{79}, 023502 (2009).

\bibitem{ref:ISW5}  D. Sapone and M. Kunz, Phys. Rev. D \textbf{80}, 083519 (2009).

\bibitem{ref:ISW6}  Y. T. Wang, L. X. Xu, and Y. X. Gui, Phys. Rev. D \textbf{82}, 083522 (2010).

\bibitem{ref:perturbationreview1} J. M. Bardeen, Phys. Rev. D {\bf 22}, 1882
(1980).

\bibitem{ref:perturbationreview2} H. Kodama, and M. Sasaki, Prog.
Theo. Phys. Supp. {\bf 78}, 1 (1984).

\bibitem{ref:perturbationreview3} C. P. Ma and E. Bertschinger, Astrophys. J.
\textbf{455}, 7 (1995).

\bibitem{ref:perturbationreview4} W. Hu and D. J. Eisenstein, Phys. Rev. D {\bf 59},
083509 (1999).

\bibitem{ref:perturbationreview5} K. A. Malik and D. Wands, Phys. Rept. {\bf
475}, 1 (2009).


\bibitem{ref:MCMCweb} http://cosmologist.info/cosmomc/.

\bibitem{ref:MCMC} A. Lewis and S. Bridle, Phys. Rev. D {\bf 66}, 103511
(2002).

\bibitem{ref:cambweb} http://camb.info/.

\bibitem{ref:noqunitom} Ch. Yeche, A. Ealet, A. Refregier, C. Tao, A. Tilquin, J.-M.
Virey, and D. Yvon, astro-ph/0507170;

P. S. Corasaniti, M. Kunz, D. Parkinson, E. J. Copeland, and B. A.
Bassett, Phys. Rev. D {\bf 70}, 083006 (2004).


\bibitem{ref:bbn} S. Burles, K. M. Nollett, and M. S. Turner, Astrophys. J. {\bf 552}, L1
(2001).


\bibitem{ref:0905}  A. G. Riess {\it et al.}, Astrophys. J. {\bf 699}, 539 (2009), arXiv:0905.0695.


\bibitem{wmapdata7} http://lambda.gsfc.nasa.gov/product/map/current/.


\bibitem{ref:Percival2} W. J. Percival {\it et al.}, Mon. Not. Roy. Astron. Soc. {\bf 401}, 2148 (2010), arXiv:0907.1660.

\bibitem{ref:werror1} W. H. Press {\it et al.}, Numerical Recipes, Cambridge University Press (1994).

\bibitem{ref:werror2} U. Alam, V. Sahni, T. D. Saini, and A. A. Starobinsky, astro-ph/0406672.

\bibitem{ref:werror3} S. Nesseris and L. Perivolaropoulos, Phys. Rev. D {\bf 72}, 123519 (2005), astro-ph/0511040.
\bibitem{ref:margnalized} T. D. Kitching and A. Amara, arXiv:0905.3383.

\end{thebibliography}
\end{document}